\documentclass[aps,prb,onecolumn,superscriptaddress,amsmath,amssymb,revsymb,showpacs]{revtex4-2}

\usepackage{color}
\usepackage{graphicx}% Include figure files
\usepackage{dcolumn}% Align table columns on decimal point
\usepackage{bm}% bold math
\usepackage{url}
\usepackage{here}
\usepackage{color}
\usepackage{CJKutf8}
\usepackage{comment}
\newcommand{\be}{\begin{eqnarray}}
\newcommand{\ee}{\end{eqnarray}}

\newcommand{\lb}{\label}
\begin{document}

%\preprint{}

\title{Spin Neural Network Potential for Magnetic Phase Transitions in Uranium Dioxide}

\author{Keita Kobayashi}%
\email{kobayashi.keita@jaea.go.jp}
\affiliation{%
CCSE, Japan Atomic Energy Agency, 178-4-4, Wakashiba , Kashiwa, Chiba
 277-0871, Japan}%

\author{Hiroki Nakamura}%
\affiliation{%
CCSE, Japan Atomic Energy Agency, 178-4-4, Wakashiba , Kashiwa, Chiba
 277-0871, Japan}%

\author{Mitsuhiro Itakura}%
\affiliation{%
CCSE, Japan Atomic Energy Agency, 178-4-4, Wakashiba , Kashiwa, Chiba
 277-0871, Japan}%
  
%\author{Masahiko Machida}%
%\affiliation{%
%CCSE, Japan Atomic Energy Agency, 178-4-4, Wakashiba , Kashiwa, Chiba
% 277-0871, Japan}%

\date{\today}% It is always \today, today,

\begin{abstract}
Uranium dioxide (UO$_2$) is a prototypical nuclear fuel material, yet predicting its thermophysical properties across a wide temperature range remains challenging.
One factor contributing to this difficulty is the complex magnetic ordering at low temperatures, where spin--orbit coupling produces strong coupling between spin and lattice degrees of freedom.
Direct DFT simulations of magnetic phase transitions at finite temperatures are computationally prohibitive. 
Here, we develop a spin neural network potential (SpinNNP) that explicitly incorporates spin degrees of freedom together with spin--orbit coupling to describe the magnetic states of UO$_2$. 
Reference datasets were generated using magnetic constrained DFT+$U$ calculations with spin--orbit coupling, covering a wide range of non-collinear spin configurations.
The SpinNNP accurately reproduces DFT energies, atomic forces, spin forces, and lattice constants. 
Machine-learning molecular dynamics simulations with spin dynamics successfully capture the antiferromagnetic-paramagnetic transition. 
Although the predicted magnetic ground state differs from experiment due to known limitations of the underlying DFT description, the transition temperature obtained is of the correct order of magnitude compared with experiment. 
These results demonstrate that machine-learning potentials can enable large-scale spin–lattice simulations of actinide oxides and provide a practical route toward predictive modeling of complex magnetic materials.
\end{abstract}

\maketitle

% \linenumbers

\section{Introduction}
Uranium dioxide (UO$_2$) is the most widely used nuclear fuel in current commercial reactors, 
and its performance is governed by a complex interplay between structural, thermal, and magnetic properties. 
In particular, UO$_2$ exhibits non-trivial magnetic ordering at low temperatures, 
where spin--orbit coupling (SOC) induces strong interactions between spin and lattice degrees of freedom. 
These magnetic effects are considered to significantly influence thermal properties such as 
thermal conductivity~\cite{moore1971thermal,gofryk2014anisotropic,liu2016molecular}, which are central to modeling fuel performance. 
Nevertheless, conventional fuel performance models still rely on empirical correlations, 
and a predictive, atomistic-level understanding of these phenomena remains incomplete.

Molecular dynamics (MD) simulations provide a powerful tool for exploring such complex behavior 
and predicting thermal properties of nuclear fuel materials. 
The reliability of MD simulations depends critically on the accuracy of the underlying interatomic potential. 
MD with classical force fields (CFFs) has been widely adopted, 
as it enables statistically converged properties to be obtained at reasonable computational cost~\cite{potashnikov2011high, CRG, cooper2014thermophysical, galvin2016thermophysical}. 
Indeed, several studies have investigated the thermal properties of UO$_2$ using classical MD. 
However, CFFs that explicitly incorporate spin degrees of freedom 
have been developed in limited cases, and none have yet been established for actinide oxides such as UO$_2$.  
An alternative approach is first-principles molecular dynamics (FPMD) based on density functional theory (DFT)~\cite{f-electon,szpunar2016theoretical, nakamura2016high}. 
While FPMD generally provides reliable predictions of material properties, 
it is computationally prohibitive for large-scale simulations, 
particularly when non-collinear magnetism and SOC are explicitly included.

In the last decade, machine learning potentials (MLPs) have emerged as a promising approach 
to bridge the gap between efficiency and accuracy.  
In these methods, flexible functional forms with multiple adjustable parameters, 
such as artificial neural networks \cite{BPNN1,BPNN2} and Gaussian processes \cite{GAP1,GAP2}, 
are trained on first-principles datasets.  
Machine learning molecular dynamics (MLMD) based on such potentials has been extensively applied to diverse systems~\cite{kobayashi2023machine, nagai2024high,urata2024applications}, 
including nuclear fuels~\cite{kobayashi2022machine, dubois2024atomistic, stippell2024building,konashi2025neural,kobayashi2025specific}. 
More recently, efforts have been made to extend MLPs by incorporating spin degrees of freedom.  
For example, collinear spin has been included in extensions of Behler--Parrinello type neural network potentials (BPNNP)~\cite{eckhoff2021high, novikov2022magnetic} 
and the momentum tensor potential~\cite{eckhoff2021high, novikov2022magnetic}.  
In addition, approaches for non-collinear spin degrees of freedom have been proposed, 
including graph neural network potentials~\cite{yu2024spin}, Gaussian process regression frameworks~\cite{gao2024machine}, and atomic cluster expansion~\cite{rinaldi2024non}.
These developments represent important steps toward accurate MD simulations with explicit spin dynamics.  
However, to date, no such potentials have been developed for actinide oxides such as UO$_2$.

In this study, we present a spin neural network potential (SpinNNP) for UO$_2$,  
which explicitly incorporates non-collinear spin degrees of freedom and SOC effects  
within the Behler--Parrinello neural network framework.  
Trained on a dataset of DFT+U calculations including non-collinear spin configurations,  
SpinNNP accurately reproduces energies, forces, spin forces, and lattice parameters.  
We further validate its capability to describe magnetic phase transitions at finite temperatures  
through machine learning molecular dynamics (MLMD) simulations,  
demonstrating its potential for large-scale simulations of actinide nuclear fuels  
with strong spin--lattice coupling.

\section{Computational Framework}
The BPNNP has been widely used to construct accurate and transferable interatomic potentials from first-principles data \cite{BPNN1,BPNN2}. 
In this framework, the total energy of the system is expressed as a sum of atomic contributions, 
$
E_{\rm NNP} = \sum_{i} e \left( \mathbf{G}_i(\{\mathbf{r}_{i}\}) \right)
$, 
where $e$ denotes the neural network associated with atom $i$, and $\mathbf{G}_i$ is a set of descriptors that encode the local atomic environment around atom $i$. 
In BPNNP, symmetry functions are employed as descriptors of interatomic distances and angles, defined as
\begin{align}
&G_{i}^{\rm (2)} = \sum_{j(\neq i)} e^{-\eta(r_{ij}-R_s)^2} f_c(r_{ij}), \\
&G_{i}^{\rm (3)} = 2^{1-\zeta} \sum_{j,k \neq i, \,(j<k)} 
\xi^{\rm (A)}(\mathbf{r}_{ij}, \mathbf{r}_{ik}), \\
&\xi^{\rm (A)}(\mathbf{r}_{ij}, \mathbf{r}_{ik}) = 
\left( 1 + \lambda \cos \theta_{ijk} \right)^{\zeta }
e^{-\eta(r_{ij}^2 + r_{ik}^2 + r_{jk}^2)}  
f_c(r_{ij})
f_c(r_{ik})
f_c(r_{jk}) \, ,
\end{align}
with the cutoff function $f_c(r_{ij})$ defined within a certain distance $R_{c}$. 
The use of symmetry functions ensures invariance of the potential with respect to translation, rotation, and permutation of chemically equivalent atoms, which is essential for transferability. 
Once the reference dataset is prepared, the neural networks are trained to reproduce the energies and forces obtained from DFT calculations. 

In addition to atomic motion, describing spin dynamics requires an interatomic potential that depends on both atomic positions and spin variables. 
In this study, we construct an MLP based on the Behler--Parrinello type neural network, expressed as
\begin{equation}
E_{\rm NNP} = \sum_{i} e \left( 
\mathbf{G}_i(\{\mathbf{r}_{i}\}, \{\mathbf{s}_{i}\}) 
\right),
\end{equation}
where $\mathbf{r}_{i}$ and $\mathbf{s}_{i}$ denote the position and spin of atom $i$, respectively. 
Here, $\mathbf{G}_i$ denotes the descriptor that encodes the local environment of each atomic center. 
To enable the potential to describe coupled spin–lattice physics, it is necessary to introduce descriptors that explicitly include spin degrees of freedom.
Subsequently, reference datasets are generated from DFT calculations performed for various atomic and spin configurations, and these data are used to train the spin neural network potential (SpinNNP). 
In this study, we modified the N2P2 code \cite{n2p2} to enable the construction of machine learning potentials that depend simultaneously on atomic positions and spins. 
The details of this framework are described in the following.

\subsection{Spin Symmetry Functions}
The simplest descriptor is the spin magnitude of each atom,
\begin{equation}
G_i^{\rm (1S)} = |\mathbf{s}_i|^2 \, ,
\end{equation}
which gives rise to a Ginzburg–Landau expansion \cite{uhl1996exchange, rosengaard1997finite, wysocki2008thermodynamics, lavrentiev2010magnetic, lavrentiev2011noncollinear} of the form 
$
e(|\mathbf{s}_i|^2) \simeq \sum_{n}a_n |\mathbf{s}_i|^{2n} .
$
As a simple form of spin–spin interaction, symmetry functions incorporating an isotropic Heisenberg-type interaction, $\mathbf{s}_i^{T} \mathbf{s}_j$, can be introduced as
\begin{equation}
G_{i}^{\rm (2S)} = \sum_{j(\neq i)} e^{-\eta(r_{ij}-R_s)^2} f_c(r_{ij}) \mathbf{s}_{i}^T\mathbf{s}_{j}\,,
\end{equation}
which we refer to as the Heisenberg-type radial symmetry function.
Here, indices $i$ and $j$ denote magnetic atoms (e.g., uranium atoms in UO$_2$).
As an extension of $G_{i}^{\rm (2S)}$ that incorporates angular dependence, we introduce
\begin{align}
G_{i}^{\rm (3S)} = 2^{1-\zeta} \sum_{j,k \neq i, \,(j<k)} 
\xi^{\rm (A)}(\mathbf{r}_{ij}, \mathbf{r}_{ik})
\mathbf{s}_j^{T}  \mathbf{s}_k
\,,
\end{align}
which we refer to as the Heisenberg-type angular symmetry function.

The spin symmetry functions introduced above are invariant under spin rotations.
However, in UO$_2$, strong spin-orbit interaction couples spins to the lattice and breaks spin rotational symmetry.
In UO$_2$, due to the strong spin--orbit interaction, spins are coupled to the lattice and thus lose spin rotational symmetry. 
However, they remain invariant under simultaneous rotation of both atomic coordinates and spins. 
In the following, we introduce symmetry functions that incorporate anisotropic spin interactions to describe the coupling between spins and the lattice. 
To capture this spin--lattice coupling, a natural starting point is to introduce anisotropic interactions in which the spin variables are coupled to the bond orientation vector $\mathbf{e}_{ij}$. 
This construction ensures invariance under simultaneous global rotations of atomic coordinates and spins.
As a natural starting point, we consider the spin interactions
\[
\mathbf{s}_{j}^{T}\mathbf{J}_{ij}\mathbf{s}_{j}, \quad
\mathbf{s}_{i}^{T}\mathbf{J}_{ij}\mathbf{s}_{i}, \quad
\mathbf{s}_{i}^{T}\mathbf{J}_{ij}\mathbf{s}_{j},
\]
where $\mathbf{J}_{ij} = \mathbf{e}_{ij}\mathbf{e}_{ij}^{T}$ is a projection operator onto the bond direction, and
$\mathbf{e}_{ij}$ is the unit vector pointing from atom $i$ to atom $j$, $\mathbf{e}_{ij} = \mathbf{r}_{ij}/r_{ij}$.
We then define the following symmetry functions as
\begin{align}
G_{i}^{\rm (2SOa)} &= \sum_{j(\neq i)} e^{-\eta(r_{ij}-R_s)^2} f_c(r_{ij}) \mathbf{s}_{j}^{T}\mathbf{J}_{ij}\mathbf{s}_{j}, \\
G_{i}^{\rm (2SOb)} &= \sum_{j(\neq i)} e^{-\eta(r_{ij}-R_s)^2} f_c(r_{ij}) \mathbf{s}_{i}^{T}\mathbf{J}_{ij}\mathbf{s}_{i}, \\
G_{i}^{\rm (2SOc)} &= \sum_{j(\neq i)} e^{-\eta(r_{ij}-R_s)^2} f_c(r_{ij}) \mathbf{s}_{i}^{T}\mathbf{J}_{ij}\mathbf{s}_{j},
\end{align}
which we denote as spin–orbit coupled radial symmetry functions.
$G_{i}^{\rm (2SOa)}$ and $G_{i}^{\rm (2SOb)}$ introduce environment-dependent uniaxial spin anisotropy by coupling the spin orientation to the local bond direction. 
For example, in UO$_2$, contributions from U–O bonds can favor spin alignment along specific bond directions, thereby inducing local anisotropy consistent with spin–orbit coupling.
In contrast, $G_{i}^{\rm (2SOc)}$ describes anisotropic spin–spin interactions projected onto the bond direction and effectively introduces an Ising-like coupling along each bond.
We also consider spin interactions of the form
\begin{equation}
\mathbf{s}_{j}^{T} \mathbf{J}_{ijk}^{\rm (sym)} \mathbf{s}_{k} \,, \quad  
\mathbf{s}_{j}^{T} \mathbf{J}_{ijk}^{\rm (asym)} \mathbf{s}_{k}\,,
\end{equation}
where
\begin{align}
\mathbf{J}_{ijk}^{\rm (sym)} &= \mathbf{e}_{ij}\mathbf{e}_{ik}^{T} + \mathbf{e}_{ik}\mathbf{e}_{ij}^{T}, \\
\mathbf{J}_{ijk}^{\rm (asym)} &= \mathbf{e}_{ij}\mathbf{e}_{ik}^{T} - \mathbf{e}_{ik}\mathbf{e}_{ij}^{T}.
\end{align}
The antisymmetric interaction matrix $\mathbf{J}_{ijk}^{\rm (asym)}$ corresponds to the Dzyaloshinskii--Moriya (DM) interaction \cite{moriya1960anisotropic, keffer1962moriya, cheong2007multiferroics}, 
\[
H_{\rm DM} = \mathbf{D}_{jk}\cdot (\mathbf{s}_{j}\times\mathbf{s}_{k}),
\]
which can be written as 
\[
\mathbf{D}_{jk}\cdot (\mathbf{s}_{j}\times\mathbf{s}_{k}) 
\propto (\mathbf{e}_{ij}\times\mathbf{e}_{ik}) \cdot (\mathbf{s}_{j}\times\mathbf{s}_{k})
= \mathbf{s}_{j}^{T} \mathbf{J}_{ijk}^{\rm (asym)}\mathbf{s}_{k}.
\]
This interaction arises from second-order perturbation theory of the spin–orbit coupling term $\mathbf{l}\cdot\mathbf{s}$ and is mediated by super-exchange between magnetic cations.
In the case of UO$_2$, the $j$-th and $k$-th atoms correspond to uranium ions, while the $i$-th atom corresponds to an oxygen ion that mediates the super-exchange pathway.
In contrast, the symmetric interaction matrix $\mathbf{J}_{ijk}^{\rm (sym)}$ is expected to describe 
anisotropic spin interactions such as those appearing in the Ising, XY, and XXZ models. 
We then introduce the following symmetry functions, 
\begin{align}
G_{i}^{\rm (3SOa)} &= 2^{1-\zeta} \sum_{j,k\neq i, \, (j<k)} 
\xi^{\rm (A)}(\mathbf{r}_{ij},\mathbf{r}_{ik})
\left(
\mathbf{s}_j^{T}
\mathbf{J}_{ijk}^{\rm (sym)}
\mathbf{s}_k
\right), \\
G_{i}^{\rm (3SOb)} &= 2^{1-\zeta} \sum_{j,k\neq i, \, (j<k)} 
\xi^{\rm (A)}(\mathbf{r}_{ij},\mathbf{r}_{ik})
\left(
\mathbf{s}_j^{T}
\mathbf{J}_{ijk}^{\rm (asym)}
\mathbf{s}_k
\right),
\end{align}
as spin–orbit coupled angular symmetry functions.

We employed the conventional symmetry functions $G_{i}^{\rm (2)}$ and $G_{i}^{\rm (3)}$, together with the spin-dependent symmetry functions introduced above, as the input descriptors of SpinNNP for UO$_2$. 
This combination enables the potential to simultaneously describe atomic structure, magnetic ordering, and spin–lattice coupling within a unified descriptor framework.
The cutoff radii of the cutoff functions $f_c(r_{ij})$ were set to 6.5~\AA\ for the radial symmetry functions and 4.5~\AA\ for the angular symmetry functions. 
In this study, we employ the polynomial cutoff function~\cite{n2p2},
\begin{equation}
f_c(x)=
\begin{cases}
1 - 6x^{5} + 15x^{4} - 10x^{3} & (x \le 1),\\
0 & (x > 1),
\end{cases}
\end{equation}
where $x = r/R_c$ is the normalized interatomic distance.
The parameters of the symmetry functions ($\eta$, $R_s$, $\zeta$, and $\lambda$) were determined via CUR decomposition \cite{CUR}. The descriptor dimensionality was set to 80 for both oxygen and uranium atoms.

\subsection{Reference Dataset Preparation and Model Fitting}

We conducted DFT calculations using the Vienna \textit{Ab-initio} Simulation Package (VASP)~\cite{VASP1,VASP2}. 
The GGA-PBEsol exchange–correlation functional~\cite{perdew2008restoring,pegg2017dft+} with a Hubbard $U$ correction ($U=3.5$~eV) and spin–orbit coupling was employed in this study. 
In all calculations, the projector augmented-wave (PAW) method~\cite{PAW} was employed, with a plane-wave energy cutoff of 400~eV and a $k$-point density of 0.5~\AA$^{-1}$. 
To generate various spin configurations, we employed the magnetic constrained DFT (cDFT) method with the constraint term
\begin{equation}
E_{\rm const} = \lambda \sum_{i} \left( \mathbf{s}_{i}^{0}- \mathbf{s}_{i} \right)^2 ,
\end{equation}
where $\mathbf{s}_{i}^{0}$ is the target spin vector and $\mathbf{s}_{i}$ is the spin vector obtained in the calculation, defined as
\begin{equation}
\mathbf{s}_{i} = \int_{\Omega_i} d\mathbf{x}\, \mathbf{s}(\mathbf{x}) F_{I}(|\mathbf{x}|),
\end{equation}
with $\Omega_i$ a sphere of radius 1.42~\AA\ centered on atom $i$ and $F_I$ a smooth decaying weight function. 
Because of the projection, the magnitude of $\mathbf{s}_{i}$ differs from the physical magnetic moment of spin,
$
\mathbf{m}_{i} = \int_{\Omega_i} d\mathbf{x}\, \mathbf{s}(\mathbf{x}) .
$
Nevertheless, $\mathbf{s}_{i}$ and $\mathbf{m}_{i}$ are nearly proportional. 
From a linear fit to the reference dataset, we obtain $\mathbf{s}_{i} = 0.6399\,\mathbf{m}_{i}$.

We randomly deformed the cell shape and atomic positions, and generated various spin configurations for both the unit cell (12 atoms) and the $2\times2\times2$ supercell (96 atoms) of UO$_2$.  
Energies and forces for these configurations were evaluated using cDFT.  
The resulting quantities are given as 
$E_{\rm cDFT} = E_{\rm DFT} + E_{\rm const}$ and 
$F_{\rm cDFT}=-\frac{\partial E_{\rm cDFT}}{\partial \mathbf{r}}$.  
Since the constraint term is an artificial contribution introduced solely to fix the spin orientations, 
the true quantities required for training are $E_{\rm DFT}$ and $F_{\rm DFT}$.  
However, $F_{\rm cDFT}$ and $F_{\rm DFT}$ are generally not identical, 
and the combination of $E_{\rm DFT}$ and $F_{\rm cDFT}$ is inconsistent with the Hellmann–Feynman theorem.  
To minimize this inconsistency, we selected only those cDFT calculations in which the constraint energy satisfied $E_{\rm const}/{\rm atom} \leq 0.01$~eV.  
Furthermore, using the cDFT wavefunctions, we evaluated the forces at $\lambda=0$ to approximate the unconstrained DFT forces, 
$
\mathbf{F}_{{\rm DFT},i}=\langle\Psi_{\rm cDFT}|
\frac{\partial \hat{H}_{\rm DFT}}{\partial \mathbf{r}_i}
|\Psi_{\rm cDFT}\rangle
$
and selected the data satisfying 
$\max \{|\mathbf{F}_{{\rm DFT},i} - \mathbf{F}_{{\rm cDFT},i}|\} \leq 0.01$~eV/\AA.  
In this way, we ensured that the adopted cDFT results, $\{E_{\rm DFT}, \{\mathbf{F}_{{\rm DFT},i}\}\}$, were approximately consistent with the Hellmann–Feynman theorem. 
We verified that these thresholds are sufficient to guarantee reliable energies and forces for training the SpinNNP.

In addition to energies and forces, we also included approximate spin forces derived from the constraint term.  
From the stationarity condition in cDFT, 
$
\frac{\partial E_{\rm cDFT}}{\partial \mathbf{s}_i} \simeq 0 
$,
we obtain an approximate relation for the spin forces, 
\[
\mathbf{B}_{{\rm DFT},i} = -\frac{\partial E_{\rm DFT}}{\partial \mathbf{s}_i} 
\simeq \frac{\partial E_{\rm const}}{\partial \mathbf{s}_i}.
\]
These quantities were incorporated as additional training targets in the dataset.  
Thus, SpinNNP was trained using the set of target values, 
$\{E_{\rm DFT}, \{\mathbf{F}_{{\rm DFT},i}\}, \{\mathbf{B}_{{\rm DFT},i}\}\}$.

Furthermore, additional DFT reference data were generated by performing MLMD simulations from 1 K to 1000K  and sampling various structures. 
NPT simulations with a Nosè--Hoover thermostat and a Parrinello--Rahman barostat were conducted to sample diverse configurations and volumes. 
The spin dynamics was described by the generalized Langevin spin dynamics (GLSD) equation\cite{ma2012longitudinal},  
\begin{align}
\frac{d\mathbf{s}_{i}}{dt} &=
\frac{1}{\hbar}
\left(
  \mathbf{s}_{i}
 \times
 \mathbf{B}_{{\rm NNP},i} 
\right) 
-\frac{\lambda s_i^2 }{\hbar} \mathbf{B}_{{\rm NNP},i} 
+\boldsymbol{\xi}_i  , \lb{eq:GLSD} 
\end{align}
where $\mathbf{B}_{{\rm NNP},i} =  -\frac{\partial E_{\rm NNP}}{\partial \mathbf{s}_i}$ is the spin force derived from the SpinNNP,  
$\lambda$ is a damping constant, and $\boldsymbol{\xi}_i$ denotes the stochastic noise term.  
The noise satisfies the statistical properties
\begin{align}
\langle \boldsymbol{\xi}_i \rangle &= 0 , \\
\langle \xi_{i\mu}(t)\, \xi_{j\nu}(t') \rangle &= D_i\delta_{ij}\,\delta_{\mu\nu}\,\delta(t-t')\,,
\end{align}
with $\mu,\nu$ denoting Cartesian components. 
The noise strength $D$ is related to the temperature through the fluctuation--dissipation theorem as
$
D_i = 2\lambda  s_{i} k_{\rm B}T/\hbar\, .
$
All MLMD simulations were performed using the LAMMPS code~\cite{Lammps}. 
To enable the present MLMD simulations with SpinNNP, we made modifications to the N2P2--LAMMPS interface~\cite{n2p2} and to the LAMMPS spin package~\cite{tranchida2018massively}.
From the structures generated by MLMD simulations, we picked up those with large variances in energy, force, and spin force as indicators of model uncertainty. 
The variances were estimated using the bootstrap method, in which five SpinNNP models were trained by dividing the training dataset. 
In addition, we selected structures located in the extrapolation region, determined by whether their symmetry function values extended beyond the minimum and maximum bounds observed in the reference dataset. 
The energies, forces, and spin forces of the selected structures were evaluated by DFT calculations to augment the training dataset. 

In total, 625 DFT configurations were used as the reference dataset.
For the evaluation, the dataset was randomly divided into 95\% training data and 5\% test data.
The neural network architecture consists of two hidden layers with 10 nodes in each layer and hyperbolic tangent activation functions.
The RMSEs of the present SpinNNP, with and without spin force fitting, are summarized in Table~\ref{tab:table1}.
In both cases, the RMSEs of the energies are below 1~meV/atom, which is comparable to the typical accuracy of modern machine-learning interatomic potentials.
Furthermore, spin force fitting reduced the RMSE of the spin forces and was also found to mitigate overfitting of the atomic forces. 
In this study, we therefore adopted the SpinNNP trained with spin force fitting.

\begin{table}[ht]
\small
\centering
\caption{Root-mean-square errors (RMSEs) of SpinNNP with and without spin force fitting. 
The dataset was split into 95\% for training and 5\% for testing.}
\label{tab:table1}
\begin{tabular*}{1\textwidth}{@{\extracolsep{\fill}}lcc}
\hline
\hline
 & Train (95\%) & Test (5\%) \\
\hline
\multicolumn{3}{l}{\textbf{SpinNNP without spin force fitting}} \\
Energy (meV/atom)          & 0.736   & 0.554  \\
Force (meV/\AA)            & 26.658  & 41.424 \\
Spin-force (meV/$\mu_{\rm B}$) & 12.106 & 12.608 \\
\hline
\multicolumn{3}{l}{\textbf{SpinNNP with spin force fitting}} \\
Energy (meV/atom)          & 0.733   & 0.560  \\
Force (meV/\AA)            & 31.136  & 32.928 \\
Spin force (meV/$\mu_{\rm B}$) & 4.885  & 5.612  \\
\hline
\end{tabular*}
\end{table}

\section{Results and Discussion} 
Having established the computational framework, we now assess the performance of SpinNNP. 
The ordered magnetic phases of UO$_2$ provide a stringent benchmark, 
owing to their complex multi-$k$ antiferromagnetic structures and the strong coupling between lattice and spin degrees of freedom. 
These characteristics allow us to evaluate not only the accuracy of SpinNNP in reproducing static DFT reference data, but also its ability to capture the subtle interplay between structural and magnetic properties.
We begin by examining the energetics and lattice parameters of various antiferromagnetic and ferromagnetic states, comparing the SpinNNP predictions directly with DFT results. 
Subsequently, we extend the analysis to finite-temperature simulations to investigate whether SpinNNP can also reproduce magnetic phase transitions. 

Regarding antiferromagnetic order, calculations were carried out for both longitudinal and transverse multi-$k$ antiferromagnetic states.
The longitudinal one-, double-, and triple-$k$ (L1k, L2k, and L3k) antiferromagnetic states are defined as  
\begin{equation}
\mathbf{s}_{i}^{\rm (Lmk)} \propto
\sum_{l=1}^{m} 
\bar{\mathbf{a}}_l \cos\!\left( \mathbf{k}_l \cdot \mathbf{r}_i \right) ,
\end{equation}
where $\mathbf{r}_i$ denotes the position vector of the $i$-th magnetic ion,  
and $\bar{\mathbf{a}}_i = \mathbf{a}_i/|\mathbf{a}_i|$ is the unit vector along the lattice vector $\mathbf{a}_i$.  
Here, the reciprocal lattice vectors are defined in general form as  
$\mathbf{k}_i = 2\pi (\mathbf{a}_j \times \mathbf{a}_k)/[\mathbf{a}_i \cdot (\mathbf{a}_j \times \mathbf{a}_k)]$,  
with $(i,j,k)$ being a cyclic permutation of $(1,2,3)$ and $\mathbf{a}_i$ the real-space lattice vectors.  
Similarly, the transverse one-$k$ (T1k), double-$k$ (T2k), and triple-$k$ (T3k) antiferromagnetic states are defined as  
\begin{align}
\mathbf{s}_{i}^{\rm (T1k)} &\propto \bar{\mathbf{a}}_1 \cos( \mathbf{k}_3 \cdot \mathbf{r}_i ) , \\  
\mathbf{s}_{i}^{\rm (T2k)} &\propto 
\bar{\mathbf{a}}_2 \cos( \mathbf{k}_1 \cdot \mathbf{r}_i ) +
\bar{\mathbf{a}}_3 \cos( \mathbf{k}_2 \cdot \mathbf{r}_i ) , \\  
\mathbf{s}_{i}^{\rm (T3k)} &\propto 
\bar{\mathbf{a}}_1 \cos( \mathbf{k}_3 \cdot \mathbf{r}_i ) +
\bar{\mathbf{a}}_2 \cos( \mathbf{k}_1 \cdot \mathbf{r}_i ) +
\bar{\mathbf{a}}_3 \cos( \mathbf{k}_2 \cdot \mathbf{r}_i ) .  
\end{align}
Ferromagnetic states with magnetization along the [100], [110], and [111] crystallographic directions 
(referred to as FM$_{100}$, FM$_{110}$, and FM$_{111}$, respectively) were also investigated.

\begin{figure}[ht]
\begin{center}
\includegraphics[width=1\linewidth]{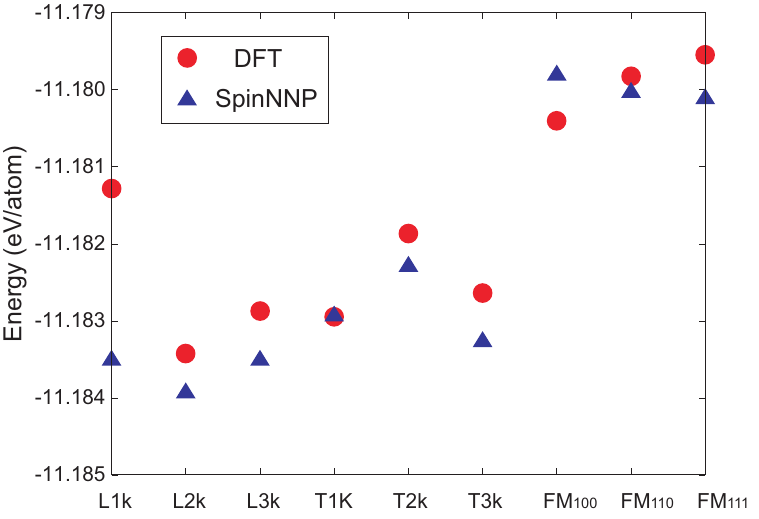}
\end{center}
\caption{
Energies for various magnetically ordered states computed by DFT and SpinNNP. 
L1k, L2k, and L3k denote longitudinal 1--3k antiferromagnetic ordered states, respectively. 
T1k, T2k, and T3k denote transverse 1--3k antiferromagnetic ordered states, respectively. 
FM$_{lmk}$ denotes ferromagnetic ordered states with spin orientation along the $[lmk]$ direction.
}
\label{Fig1}
\end{figure}

Using the aforementioned antiferromagnetic and ferromagnetic states as initial configurations, 
we performed structural optimizations with respect to atomic positions, cell geometry, spin orientations, and spin magnitudes. 
In this study, the plane-wave energy cutoff in all DFT calculations was set to 400~eV, which is somewhat lower value for the DFT calculation of UO$_2$, owing to the high computational cost of non-collinear spin calculations. 
To compensate for the reduced cutoff, Pulay stress corrections were applied during the DFT structural optimizations, 
corresponding to an isotropic pressure adjustment of $-9.3$~kbar, thereby ensuring the reliability of the calculated results.

The calculated energies of the optimized magnetic states are summarized in Fig.\ref{Fig1}, 
where the results obtained from SpinNNP are compared with those from DFT. 
For the L1k antiferromagnetic state, a somewhat larger discrepancy between DFT and SpinNNP was observed; 
however, the overall trends are reproduced reasonably well. 
In both the present DFT and SpinNNP calculations, 
the L2k antiferromagnetic state was found to be the most stable. 
However, experimental studies of UO$_2$ indicate that the T3k antiferromagnetic state is the true ground state \cite{ikushima2001first}. 
The exchange–correlation functional employed in this study thus fails to reproduce the experimental ground state. 
It is  known that the realized ground state in DFT calculations strongly depends on the choice of exchange–correlation functional \cite{pegg2017dft+,zhou2022capturing}. 
Although this difficulty is inherent to $f$-electron systems in DFT, 
the present study demonstrates that SpinNNP is capable of reasonably reproducing the reference DFT results, 
even though the small energy differences between magnetic states are not fully captured.

\begin{table*}[ht]
\small
\caption{
Lattice constants ($a$, $b$, $c$ in \AA), angles ($\alpha$, $\beta$, $\gamma$ in $^\circ$), and uranium spin magnetic moments ($s$, $m$ in $\mu_\mathrm{B}$) computed by DFT and SpinNNP for various magnetic states of UO$_2$. Relative errors (in \%) with respect to DFT are also shown.
}

\label{tab:table2}
\begin{tabular*}{1.0\textwidth}{@{\extracolsep{\fill}}llcccccccc}
\hline
\hline
Order & & $a$ (\AA) & $b$ (\AA) & $c$ (\AA) & $\alpha$ ($^\circ$) & $\beta$ ($^\circ$) & $\gamma$ ($^\circ$) & $s$ ($\mu_\mathrm{B}$) & $m$ ($\mu_\mathrm{B}$) \\
\hline
       & DFT       &5.4808 &5.4782 &5.4782 &90.00 &90.00 &90.00 &0.919 &1.436 \\
L1k    & SpinNNP   &5.4877 &5.4747 &5.4747 &90.00 &90.00 &90.00 &0.880 &1.376 \\
       & Error     & (0.13) & (0.06) & (0.06) & (0.00) & (0.00) & (0.00) & (-4.21) & (-4.21) \\
\hline
       & DFT       &5.4795 &5.4795 &5.4727 &90.00 &90.00 &90.00 &0.872 &1.362 \\
L2k    & SpinNNP   &5.4798 &5.4798 &5.4723 &90.00 &90.00 &90.00 &0.877 &1.371 \\
       & Error     & (0.00) & (0.00) & (0.01) & (0.00) & (0.00) & (0.00) & (0.65) & (0.65) \\
\hline
       & DFT       &5.4771 &5.4771 &5.4771 &90.00 &90.00 &90.00 &0.868 &1.356 \\
L3k    & SpinNNP   &5.4782 &5.4782 &5.4782 &90.00 &90.00 &90.00 &0.870 &1.359 \\
       & Error     & (0.02) & (0.02) & (0.02) & (0.00) & (0.00) & (0.00) & (0.22) & (0.22) \\
\hline
       & DFT       &5.4843 &5.4738 &5.4761 &90.00 &90.00 &90.00 &0.909 &1.420 \\
T1k    & SpinNNP   &5.4870 &5.4748 &5.4745 &90.00 &90.00 &90.00 &0.892 &1.395 \\
       & Error     & (0.05) & (0.02) & (0.03) & (0.00) & (0.00) & (0.00) & (-1.79) & (-1.79) \\
\hline
       & DFT       &5.4784 &5.4784 &5.4776 &90.00 &90.00 &90.00 &0.877 &1.370 \\
T2k    & SpinNNP   &5.4812 &5.4812 &5.4743 &90.00 &90.00 &90.00 &0.873 &1.365 \\
       & Error     & (0.05) & (0.05) & (0.06) & (0.00) & (0.00) & (0.00) & (-0.40) & (-0.40) \\
\hline
       & DFT       &5.4777 &5.4777 &5.4777 &90.00 &90.00 &90.00 &0.869 &1.358 \\
T3k    & SpinNNP   &5.4790 &5.4790 &5.4790 &90.00 &90.00 &90.00 &0.899 &1.404 \\
       & Error     & (0.02) & (0.02) & (0.02) & (0.00) & (0.00) & (0.00) & (3.41) & (3.41) \\
\hline
       & DFT       &5.4898 &5.4750 &5.4750 &90.00 &90.00 &90.00 &0.935 &1.461 \\
FM$_{100}$ & SpinNNP &5.4897 &5.4746 &5.4746 &90.00 &90.00 &90.00 &0.888 &1.388 \\
       & Error     & (0.00) & (0.01) & (0.01) & (0.00) & (0.00) & (0.00) & (-4.98) & (-4.98) \\
\hline
       & DFT       &5.4829 &5.4829 &5.4742 &90.00 &90.00 &90.40 &0.879 &1.373 \\
FM$_{110}$ & SpinNNP &5.4826 &5.4826 &5.4744 &90.00 &90.00 &90.35 &0.897 &1.401 \\
       & Error     & (0.00) & (0.00) & (0.00) & (0.00) & (0.00) & (0.05) & (2.03) & (2.03) \\
\hline
       & DFT       &5.4798 &5.4798 &5.4798 &90.29 &90.29 &90.29 &0.873 &1.365 \\
FM$_{111}$ & SpinNNP &5.4800 &5.4800 &5.4800 &90.24 &90.24 &90.24 &0.899 &1.406 \\
       & Error     & (0.00) & (0.00) & (0.00) & (0.06) & (0.06) & (0.06) & (2.99) & (2.99) \\
\hline
\hline
\end{tabular*}
\end{table*}

Table~\ref{tab:table2} summarizes the lattice parameters and spin magnetic moments obtained from DFT and SpinNNP for each magnetic state.  
The maximum deviation of the lattice parameters between DFT and SpinNNP is only 0.13\% for the L1k antiferromagnetic state, whereas the largest deviation in spin magnetic moments reaches $-4.98\%$, which is somewhat larger than the lattice parameter error.  
All calculated lattice constants for the various magnetic orders are close to the experimental value of 5.4711~\AA~\cite{leinders2015accurate}.  
Although the quantitative differences in lattice constants among the magnetic states are small, qualitative distinctions in crystal symmetry can still be identified.  
Consistent with their magnetic ordering patterns, the L1k/L2k and T1k/T2k states exhibit tetragonal symmetry, whereas the L3k and T3k states retain cubic symmetry.  
Interestingly, for the ferromagnetic states, magnetization along the [100] direction preserves the tetragonal symmetry, while magnetization along the [110] and [111] directions induces a distortion of the unit cell and breaks this symmetry.  
This behavior originates from the strong spin--orbit coupling in UO$_2$, demonstrating that the SpinNNP developed in this study can reliably capture the coupling between lattice distortions and magnetization.  
Such strong coupling between magnetic ordering and lattice deformation has been experimentally reported in UO$_2$ as the magnetopiezoelectric effect~\cite{jaime2017piezomagnetism}.  
For the experimentally observed T3k magnetic order, it is known that the oxygen atoms are displaced from the ideal $8c$ position of the $Fm\bar{3}m$ structure by $\delta_{\rm O} = 0.014$~\AA \cite{faber1976neutron, ikushima2001first,desgranges2017actual}. 
In the present calculations, the oxygen displacement obtained from DFT is $\delta_{\rm O,DFT} = 0.0152$~\AA, while the SpinNNP prediction gives $\delta_{\rm O,NNP} = 0.0165$~\AA.  
These values are in good agreement with the experimental magnitude, indicating that the SpinNNP is able to reproduce subtle spin--lattice coupling effects associated with the T3k magnetic ordering.

We have shown that the present SpinNNP reasonably reproduces the energies and lattice parameters obtained from DFT calculations.
Finally, we demonstrate the capability of SpinNNP to describe magnetic phase transitions at finite temperatures.
MLMD simulations were performed by gradually heating the system from 1~K to 40~K, starting from the L2k antiferromagnetic state, which represents the ground state of the present model and was obtained through structural optimization.
The atomic motion and cell deformation were controlled by a Nos\`e--Hoover thermostat and a Parrinello--Rahman barostat, whereas the spin precession dynamics were described by the GLSD equation~(\ref{eq:GLSD}).
Simulations were carried out for a $6\times6\times6$ supercell consisting of 2592 atoms, with a time step of 0.1~fs and a damping constant of $\lambda=0.05$, for a total simulation time of 400~ps.
The physical quantities were evaluated during the heating process from 1~K to 40~K over 400~ps by averaging the trajectories and corresponding quantities within 1~ps windows.
In addition, cooling simulations from 40~K to 1~K, starting from the paramagnetic state, were also performed to examine the reversibility of the transition. 
Since experimental measurements have established that the magnetic transition in UO$_2$ is first-order~\cite{huntzicker1971magnetic}, cooling simulations from 40~K to 1~K, starting from the paramagnetic state, were also performed to examine the reversibility of the transition.

\begin{figure*}[ht]
\begin{center}
\includegraphics[width=1\linewidth]{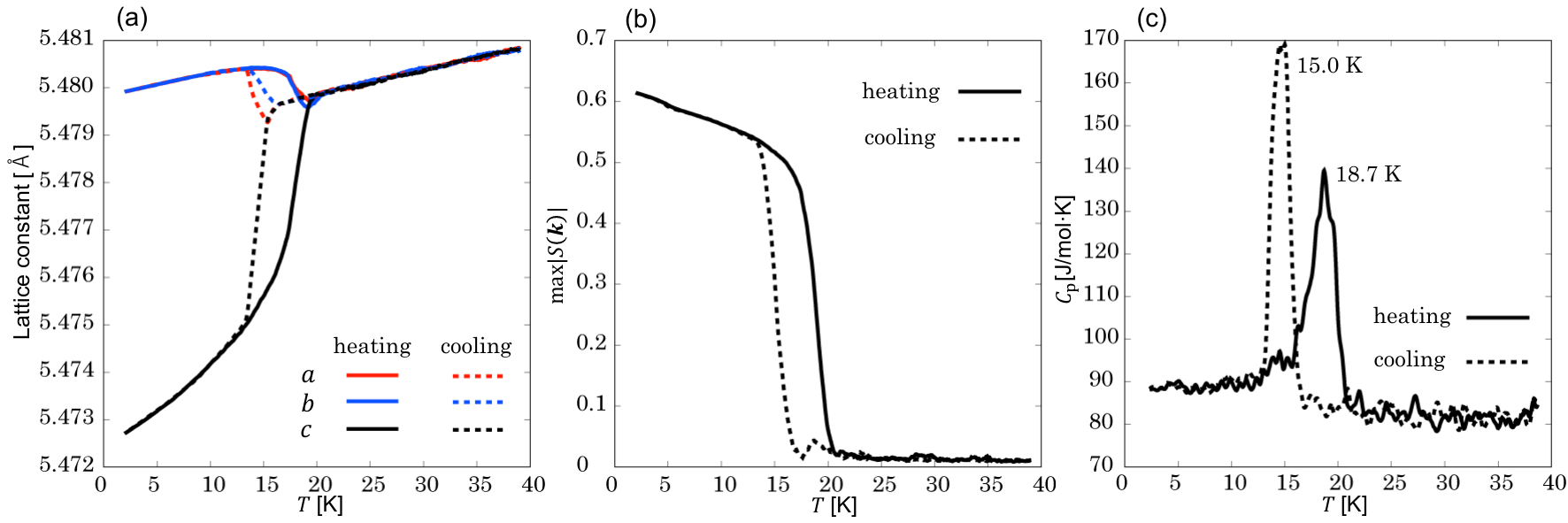}
\end{center}
\caption{
(a) Temperature dependence of the lattice constants of UO$_2$ obtained from MLMD simulations.  
(b) Order parameter $\max|S_{\mathbf{k}}|$ characterizing the magnetic ordering as a function of temperature.
(c) Calculated specific heat $C_p$ as a function of temperature, obtained from numerical differentiation of the enthalpy. 
Solid lines denote heating simulations from 1~K to 40~K, whereas dashed lines correspond to cooling simulations from 40~K to 1~K. 
The hysteresis between heating and cooling runs reflects the first-order nature of the magnetic transition.
}
\label{Fig2}
\end{figure*}

Fig.~\ref{Fig2}(a) shows the temperature dependence of the lattice constants obtained from MLMD simulations.  
At low temperatures, the relation $a=b\neq c$ reflects the tetragonal symmetry of the L2k antiferromagnetic state. 
As the temperature increases, this anisotropy disappears and the system transforms into a cubic phase, indicating a structural response to the magnetic transition.
A clear hysteresis between heating and cooling runs is observed near the transition region, consistent with the first-order character of the transition.
The discontinuous change is most pronounced along the c axis, indicating strong coupling between magnetic ordering and lattice distortion.
To clarify the magnetic phase transition, we computed the Fourier transform of the spin configuration,  
\begin{equation}
\mathbf{s}_\mathbf{k} = \frac{1}{N}\sum_{i=1}^{N} 
\mathbf{s}_{i}\, e^{-i \mathbf{k}\cdot \mathbf{r}_i} ,
\end{equation}
where $N$ is the total number of magnetic ions. 
As an order parameter, we employed $\max|\mathbf{s}_\mathbf{k}|$,  
which remains finite in magnetically ordered phases and vanishes in the paramagnetic phase, reflecting the absence of long-range order.  
The temperature dependence of the order parameter in Fig.~\ref{Fig2}(b) exhibits a discontinuous drop during heating and a corresponding recovery during cooling, demonstrating the first-order nature of the magnetic phase transition. 
The L2k antiferromagnetic order collapses upon heating and re-emerges upon cooling.

The specific heat capacity, shown in Fig.~\ref{Fig2}(c), exhibits sharp peaks at 18.7~K during heating and 15.0~K during cooling simulations, providing a clear signature of the first order magnetic phase transition.
Because the present simulations neglect quantum effects, the computed specific heat remains finite even at low temperatures, rather than approaching zero as required by quantum statistics.
Extending the present spin–lattice dynamics framework to include nuclear quantum effects, for example through path–integral molecular dynamics, would be highly challenging and lies beyond the scope of this study.
It should also be noted that the present simulations predict a magnetic ground state different from experiment and neglect nuclear quantum effects. 
Nevertheless, the predicted transition temperature (15.0–18.7~K) is lower than the experimental value of 30.44~K~\cite{huntzicker1971magnetic} but remains within a physically reasonable range, indicating that the SpinNNP framework provides a meaningful description of the transition.

\section{Conclusion}

In this study, we developed a spin neural network potential (SpinNNP) for UO$_2$ by extending the Behler--Parrinello type neural network framework to explicitly include spin degrees of freedom. 
New spin symmetry functions were introduced to describe spin-orbit coupling effects, and the potential was trained on 625 DFT configurations using energies, atomic forces, and approximate spin forces derived from constrained DFT. 
Inclusion of spin force fitting improved the accuracy of both spin and atomic forces while mitigating overfitting.
Benchmark calculations across various magnetic states demonstrated that SpinNNP accurately reproduces DFT energetics, lattice parameters, and magnetoelastic coupling, including symmetry changes induced by spin orientation. 
Finite-temperature MLMD simulations further captured the magnetic phase transition of UO$_2$, with the order parameter and specific heat indicating a transition in the range of 15-19~K, consistent in order of magnitude with the experimental value of 30.44~K.
The exchange--correlation functional employed in this study does not reproduce the experimentally observed transverse triple-$k$ ground state, instead favoring a longitudinal double-$k$ configuration. 
Identifying electronic-structure approaches capable of simultaneously describing the correct magnetic ground state and accurate structural properties remains an important challenge. 
Advanced functionals such as SCAN, hybrid methods (e.g., HSE06), or beyond-DFT approaches including DFT+DMFT may offer improved accuracy, albeit at substantially increased computational cost for non-collinear spin--orbit calculations.
In this context, machine-learning approaches such as transfer learning or fine-tuning strategies provide a promising pathway to bridge the gap between accuracy and computational efficiency. 
The present results demonstrate that SpinNNP enables large-scale spin--lattice simulations of actinide oxides and opens a route toward predictive modeling of complex magnetic materials at finite temperatures.

\section*{Declaration of Competing Interest}
The authors declare that they have no known competing financial interests or personal relationships that could have appeared to influence the work reported in this paper.

\section*{Acknowledgements}
K.K. was partially supported by JSPS KAKENHI Grants Number 24K08574 and 25K01658.  
H.N. was partially supported by JSPS KAKENHI Grant Number 23K04637.  
The calculations were mainly performed on the HPE SGI8600 supercomputing system at the Japan Atomic Energy Agency.  
We thank all staff members of CCSE for their computational support.



%% References


\bibliography{reference}

\end{document}